# Creating Contexts of Creativity:
# Musical Composition with Modular Components

Gideon D'Arcangelo
Edwin Schlossberg Incorporated
641 Sixth Avenue
New York, NY 10011 USA
+1 212 989 3993
gideon@esinter.com

## INTRODUCTION

This paper describes a series of projects that explore the possibilities of musical expression through the combination of pre-composed, interlocking, modular components. In particular, this paper presents a modular soundtrack recently composed by the author for "Currents of Creativity," a permanent interactive videowall installation at the Pope John Paul II Cultural Center which is slated to open Easter 2001 in Washington, DC.

## PRIOR RELATED WORK

### Echo

The Eloise W. Martin Center (ECHO) is an interactive music learning center permanently located at the Chicago Symphony Orchestra. It was designed by Edwin Schlossberg Incorporated (ESI) and opened in 1998; as interactive designer for the project, the author helped create software programs that sought, through various strategies, to engage visitors in the act of making music.

Visitors to ECHO select one of four custom-designed portable electronic instruments which function as the key to a series of networked interactive exhibits. Each electronic instrument is behaviorally mapped to one of four main instrument types — aerophones, chordophones, membranophones and idiophones. The instruments, which employ force-sensing resistors for input, emit an identification number that enables the system to track the visitor's path through the exhibits, and later feed back aspects of his or her performances.

The exhibits in ECHO are housed in glass booths which contain a touch screen interface and a socket on which the visitor fastens the portable instrument interface to initiate activity. There are five subject areas — *Sounds and Silence*, *Teams*, *Celebrations and Time*, *Links*, and *Mapping and Recording*. Within each subject area, there are three successive levels of activity the visitor progresses through. I will focus on the first level of *Teams*, in which a simple combinatory approach to interlocking musical components is employed.

The visitor is presented with a selection of pre-composed musical phrases associated with an instrument, such as guiro, mbira, claves, cowbell, castanets, etc. The visitor may audition these and then arrange any four of them in a stack to create an ensemble. Any of the myriad possible phrase combinations will "work" — that is, will lock together in an integrated performance. Once visitors have put together their musical teams, they are invited to improvise on their instrument along with the group.

This activity was kept very simple to make it short and focussed for our intended audience at ECHO. Nonetheless, it contains the seeds of more complex musical composition systems that offer the same interlocking architecture in more flexible environments. In such systems, if the number of possible unique combinations becomes great enough, the creative process begins to be simulated for the user, while the "safety net" of musical quality provided by pre-composed segments is still retained.

The author, along with Ben Rubin of EAR Studio in New York and Edwin Schlossberg Incorporated, successfully created such a system — a combinatory compositional environment — in the form of a robust prototype for an electronic music toy. The toy, which is still in development, provides a tangible user interface for players to combine and overlap pre-composed musical phrases to create thousands of unique arrangements.

### The Global Jukebox

The author has had many years experience working with the folklorist and ethnomusicologist Alan Lomax on the Global Jukebox, an intelligent database of world song and dance styles. The development of the Global Jukebox has been primarily funded by the National Science Foundation and Apple Computers, while also receiving funding from the Rex Foundation and Interval Research Corporation among many others. This system, which leverages one of the most comprehensive audio-visual collections of human performance style, enables users to discern large patterns of cultural dissemination. Utilizing a song style profiling system developed by Lomax and his colleagues, the Global Jukebox is able to demonstrate how any given song or group of songs fits within the matrix of world culture.

The phrases used in ECHO and subsequently in the toy prototype are highly syncopated rhythms based on musical patterns of sub-Saharan African origin. When overlapped, the phrases create polyrhythmic textures. Polyrhythmic textures are very flexible and forgiving — as long as one adheres to the overall steady pulse of the group, syncopations "work" on any beat. The inclusiveness of African polyrhythmic style is unparalleled in world music; the style permits participants of all levels of ability to play together in tight synchrony, while always leaving room for the contribution of additional layers of syncopated rhythmic pattern.

The African polyrhythmic style provides a model for the design of modular musical composition systems because it is, in a sense, already modular. It is this framework that was employed in the modular soundtrack for "Currents of Creativity."

**PROJECT DESCRIPTION**

This section describes "Currents of Creativity," an interactive exhibit designed by Edwin Schlossberg Incorporated for the Pope John Paul II Cultural Center. The author managed the interactive design team and later oversaw the production of all audio-visual media for the Center.

Located in the Gallery of Imagination, an exhibit area devoted to exploring the relationship between creativity and spirituality, "Currents" is a group compositional interactive comprised of six individual "creation" stations facing a high-resolution (2560 by 1024) videowall measuring 10' tall by 24' wide. Upon logging in to a creation station, the visitor is invited to make a visual collage based on a word of the day, such as "faith," "mystery" or "community." The videowall, meanwhile, presents a dynamic 3D environment in which background graphics and words slowly float and drift, giving the effect of a flowing, watery "current." The word of the day is prominently featured in the current at any given time.

The creation stations employ a light-box metaphor; visitors compose collages by selecting from hundreds of colorful, translucent graphics (which were created by Angela Greene, ESI's art director for interactive projects) and arranging them as they please on the light-box. When visitors complete their collages, they launch them on the wall. Their collage drifts off their screen and appears at the base of the big wall for a brief "launch" period; it is then released into the current, where it slowly glides along a path for a minute or two — interacting with the collages of other visitors before drifting off the wall.

**Modular Music Soundtrack**

There is a music soundtrack for "Currents of Creativity" which continuously bathes the Gallery of Imagination in sound. The primary design goal of the music soundtrack was to maintain a connection between visitors' actions and their perception of the music in the room. If we were not successful in making the visitor perceive his or her effect on the soundtrack, then there would no reason to make the soundtrack interactive; in that case, we could simply play a lively pre-recorded piece and the effect would be the same. It was our goal to reinforce a clear relationship between the appearance of a visitor collage and a noticeable change in the musical texture; the visitor could then take ownership of the change and associate it with their contribution.

The soundtrack is made up of the following component building blocks:

- Bed Loops. There are six of these 16-bar loops, arranged in priority order from BED01 to BED06. These provide the ambient background bed that fills the room when the program is in its default state (no collages launched).

- Launch Fanfares. There are twelve of these musical announcements, two for each creation station (there are a maximum two collage launches per station). These are designed to signal the launch of a new collage on the wall.

- Collage Loops. There are twelve of these component phrases, each played in a unique timbre. When a collage is launched, one of these twelve loops is associated with it for the duration of its life on the wall.

When there are no collages on the wall, all six tracks of the ambient bed play. When the first collage is launched, for example, BED06 fades out, a Launch Fanfare plays and a Collage Loop begins. When the second collage is launched, BED05 fades out and a second Collage Loop is triggered, and so on. Collages Loop 6 through 12 play with no underlying bed. When there are only 5 collages remaining, BED01 is reintroduced, and so on until all six bed tracks are reinstated.

This was our strategy for keeping the sound in the room percolating at all times, while reducing sound density in crowded moments so that visitors could perceive the music associated with their collage.

**Compositional Process**

The primary design challenge of the music soundtrack was to create a framework in which a visitor can trigger the launch of a Collage Loop at any moment in time. Moreover, the musical framework would have to be flexible enough to accomodate up to twelve Collage Loops simultaneously; the music would have to be perceived as an integrated whole *regardless of when the component parts were triggered.* Each loop would have to work with the other loops in myriad combinations, no matter how they happened to overlap or coincide. The result is not so much a singular composition but a flexible system that produces a range of unique performances.

After various iterations, the ambient bed for the piece was established, using polyrhythmic patterns derived from African styles. An Afro-Cuban *clave* pattern is a defining feature of the bed. Because the intention of the piece is somewhat contemplative and devotional, the polyrhythms are emphasized more softly and the syncopations are less forcefully accented than they would be in a more stylistically typical piece. Another important aesthetic consideration for the piece was its effect on museum staff who would be hearing it eight hours a day; a difficult but essential design criterion to fulfill was making the components as soothing and repeatable as possible.

*Proceedings of the CHI'01 Workshop on New Interfaces for Musical Expression (NIME-01), Seattle, USA*

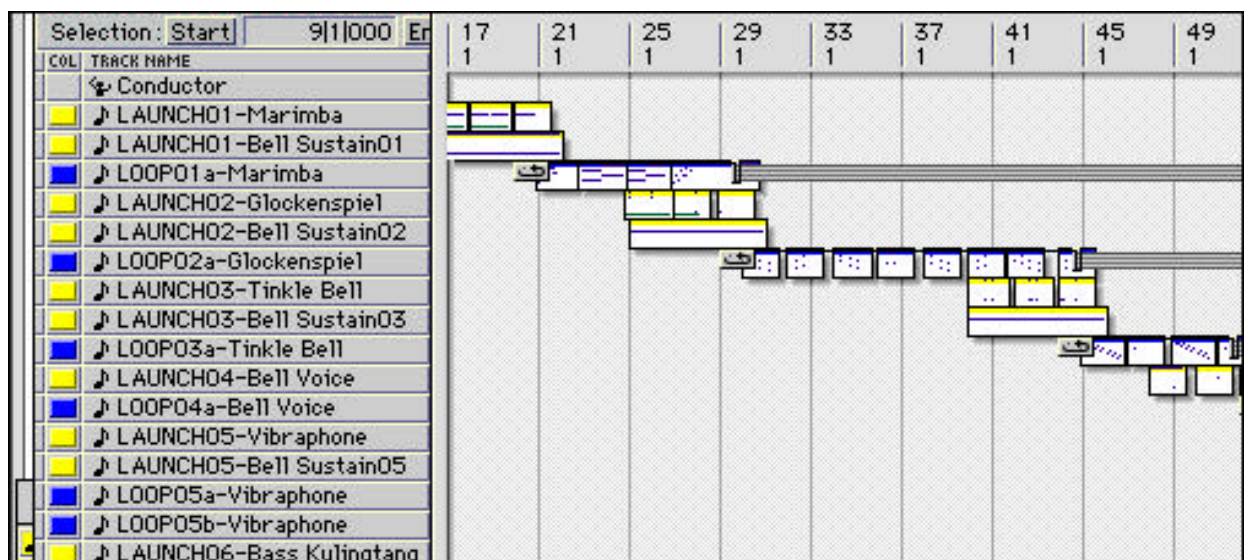

*Figure 1. Collage Loops in Digital Performer 2.7 composition environment, which allows loops to be triggered on any measure to ensure their modularity.*

The composition was completed in Digital Performer 2.7, using QuickTime Musical Instruments for initial sketches and then the *Heart of Africa* and *Heart of Asia* sample set [3,4] on an Akai S5000 for final rendering prior to recording. As loops were composed, they were checked against each other for compatibility by systematically moving their start time in Digital Performer through all possible combinations (see Figure 1). It was decided that loops could only be triggered on the first beat of a measure. The measure length in the composition is less than two seconds, which was deemed to be an acceptable response threshold for the appearance of a collage on the wall after a launch was requested.

When the piece was originally conceived, it was imagined that a fanfare, in the traditional sense of the term, would herald the arrival of a new collage on the wall. This approach proved untenable, because the highly articulated patterns of one fanfare collided with those of another when they were triggered simultaneously or within a measure of two of each other. The collisions became unavoidable when multiplied by six, so a simpler approach was adopted. A bell sample is struck to announce a new collage; depending on the station, the bell strikes one of six notes in a G-major triad over two octaves. At the same time that the bell chimes, the visitor's instrument tolls repeatedly to introduce its timbre before its Collage Loop begins; this helps anchor the visitor's musical perception to the visual display of his or her collage.

*Tonality*

The bed is intended to provide ambient background that does not interfere with the foreground Collage Loops; for this reason, non-tonal percussion instruments were employed — wet scraper, shakers, rainstick, cymbal, clave and tabla.

Tonality is a major hurdle to overcome when preparing an environment where unique combinations of pre-composed phrases can be freely staggered and overlapped. For ECHO, a pentatonic scale was employed to avoid jarring dissonances. This same approach was used for "Currents," however, not as strictly. A dominant scale was often used — G A B C D E F (F#) G — and prolonged dissonances were avoided by placing greater emphasis on pitches from the G pentatonic scale.

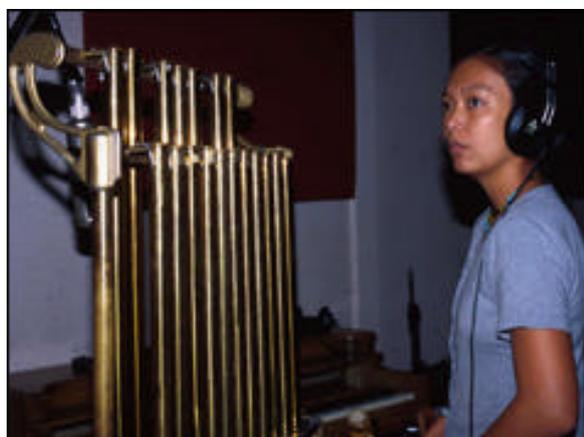

*Figure 2. Percussionist Susie Ibarra recording Orchestral Chime Collage Loop.*

**Recording Process**

We decided that it would greatly enhance the feeling and tone of the Currents soundtrack to use real instruments. The Collage and Bed Loops were performed by Susie Ibarra, a jazz percussionist and composer who has performed with the David S. Ware Quartet, John Zorn and Pauline Oliveros, among others. The Collage Loops were performed on chromatic percussion instruments —marimba, vibraphone, glockenspiel, kulingtang, alm-glocken, orchestral chimes (see Figure 2), angklung, hand bells, boo-bams (tonal drums) — and trap set.

**Implementation Process**
The software application for "Currents of Creativity" was produced by the Redmon Group of Alexandria, VA under the direction of lead programmer, Ken Cline. The 3D Display Component was created by Digital Radiance, Inc. of Madison, AL using OpenGL. The Audio Component was written in Visual Basic by Jeremy Wetzel of the Redmon Group. Ultimately, the music soundtrack was implemented as a set of modular, interlocking WAV files that are triggered by the Visual Basic application based on visitor input.

**Audio-visual Examples**
Examples of Bed Loops, Launch Fanfares, and Collage Loops, as well as examples of the fully integrated soundtrack in action, are available at fargo.itp.tsoa.nyu.edu/~publicplaces/currents.htm.

## CONCLUSION: CONTEXTS OF CREATIVITY
One fruitful direction to explore in the design of interfaces for musical expression is the creation of contexts that support creative activity. The projects discussed above begin to explore the possibilities for this area of research. By setting up universes of modular, pre-recorded, interlocking musical components — and creating a set of rules by which they may be combined — the designer has the potential for creating truly creative tools. If the rules of combination are interesting and intuitive enough, and if the volume of components is expansive enough, the number of combinations possible within the system may reach a threshold that begins to simulate the creative process.

Such tools reflect trends in popular aesthetics, where musical expression through the re-purposing, sampling, borrowing and appropriation of pre-existing recordings is recognized as a primary form of creativity. Turntablists and DJs since the 1970s have been finding evermore creative ways to outdo each other in recycling and recombining the corpus of popular music. This approach to interface design attempts to harness that creative energy.

Finally, this approach is very useful when considering the design of musical interfaces for public places such as museums, educational environments and visitor centers. When thinking of a compositional interactive for a public place, the question that always arises is "How much creative freedom do we give the visitor?" This is an area the author explores with students in his "Interactive Computing in Public Places" class in the Interactive Telecommunications Program at New York University. For example, if the exhibit is about playing the sitar, there are two extremes: we can simulate the virtuosity required to play this instrument (and thus ensure extreme difficulty and inevitable frustration for first-time users) or we can make it too simple — whereby pressing a button on a sitar-shaped device the visitor triggers a masterful raga. Since we are always dealing with first-time users in public places and the time required to learn an interface must be reduced to its minimum, it is better to find some middle ground between virtuosity and simplicity. The modular approach to creativity — providing a safety net under the creative process — is often a fruitful one to take when designing interfaces for novices that nonetheless yield satisfying experiences.

## ACKNOWLEDGMENTS
The author would like thank everyone at Edwin Schlossberg Incorporated, especially the design teams for ECHO and the Pope John Paul II Cultural Center; Ben Rubin of EAR Studio; and the Alan Lomax Collection.